\documentclass[journal, onecolumn]{IEEEtran}
\usepackage{comment}
\usepackage{amssymb}
\usepackage{amsmath}
\usepackage{amsthm}
\usepackage{amsmath,amsthm}
\usepackage{graphicx,epsfig}
\usepackage{wrapfig}

\renewcommand{\le}{\leqslant}
\renewcommand{\ge}{\geqslant}
\newtheorem{theorem}{Theorem}

\newtheorem*{lemma}{Lemma}
\newtheorem{corollary}{Corollary}

\theoremstyle{remark}
\newtheorem{remark}{Remark}

\newtheorem{definition}{Definition}
\newtheorem{example}{Example}

\usepackage{array}
\usepackage{amssymb}

\usepackage{color}

\newcommand{\prob}{\mathrm{Pr}}

\newcommand*\lon{%
       \mskip5mu
        \relax
        {:}%
        \mskip5mu
        \relax
}

\renewcommand{\mid}{|}

\newtheorem*{condition*}{}

\title{A Conditional Information Inequality  and its Combinatorial Applications}

\begin{document}

\title{A Conditional Information Inequality  and its Combinatorial Applications %
  \thanks{While working on the paper the authors were in part supported 
    by RFBR grants 14-01-93107, 16-01-00362, by an ANR-15-CE40-0016-01 grant RaCAF,
    and by the Russian Academic Excellence Project `5-100'.}
}

\author{%
Tarik Kaced\thanks{%
Tarik Kaced is with Universit\'e de Paris-Est Cr\'eteil, LACL, UPEC, France},
Andrei Romashchenko\thanks{%
Andrei Romashchenko is with LIRMM, CNRS \& Univ. Montpellier},
and
Nikolay Vereshchagin\thanks{%
Nikolay Vereshchagin is with National Research University  Higher School of Economics
}
}
%

%\IEEEpeerreviewmaketitle
\maketitle

\begin{abstract}

We show that the inequality    
$H(A \mid B,X) + H(A \mid B,Y) \le H(A\mid B)$
for jointly distributed random variables $A,B,X,Y$, which does not hold in general case,
holds under some natural condition on the support of the probability distribution of $A,B,X,Y$.
This result generalizes a version of the conditional Ingleton inequality: if for some distribution $I(X\lon Y \mid A) = H(A\mid X,Y)=0$, then $I(A\lon B) \le I(A\lon B \mid X) + I(A\lon B \mid Y)  + I(X\lon Y)$.

We present two applications of our result.
The first one is the following easy-to-formulate 
theorem on edge colorings of bipartite graphs:
assume that the edges of a bipartite graph are colored in $K$
colors so that each two edges sharing a vertex have different colors
and for each pair (left vertex $x$, right vertex $y$) 
there is at most one color $a$ such both $x$ and $y$ are incident to
edges with color $a$;
assume further that the degree of each left vertex
is at least $L$ and the degree of each right vertex is at least $R$.
Then $K\ge LR$.
The second application is a new method to prove lower bounds for biclique cover
of bipartite graphs.

%We also provide a new relaxation of the conditions for the inequality 
%proven by Z.~Zhang, R.W.~Yeung (1997).

\end{abstract}

%\begin{keywords}
%Shannon entropy; conditional information inequalities; non-Shannon-type information inequalities
%\end{keywords}

\section{Introduction}

The most general and fundamental properties of Shannon's entropy can be expressed in the language of linear inequalities.
The usual universal information inequalities (the linear inequalities that hold
for Shannon's entropies of jointly distributed tuples of random variables for
every distribution) have many equivalent characterizations and interpretations
in very different areas ---  
these inequalities can be equivalently reformulated in the settings of  Kolmogorov complexity and group theory;  
they give characterizations of  the network coding capacity rates,  of the cardinalities of projections of finite sets, etc., see the surveys in~\cite{yeung-book,facets}.
The parallel and interplay between different ``incarnations'' of information
inequalities lead to their better understanding and to more efficient applications of this technique. 
However, there exists a class of less common information inequalities that still lack a satisfactory explanation and have no clear combinatorial interpretation.
These are the \emph{conditional linear information inequalities}, which
hold only for distributions that satisfy some constraints.
% (that can be expressed as linear equations on the random variables entropies).
The first nontrivial example of a conditional linear information inequality was proven in the seminal paper \cite{zhang-conditional}; see a survey of other similar results in \cite{kaced-romashchenko-2013}.
Until now, these inequalities looked like artifacts without practical or theoretical application.
In this paper, we argue that some conditional inequalities can be naturally interpreted in purely combinatorial terms.
We propose a new ``conditional information inequality,'' discuss its
combinatorial meaning, and  show how it can be employed in purely combinatorial
proofs.
%%%%
\medskip

Let $A,X,Y$ be jointly distributed discrete random variables.
In this paper, we consider the inequality
\begin{eqnarray}\label{eq13-bis}
H(A \mid X) + H(A \mid Y) &\le& H(A),
\end{eqnarray}
where $H(\cdot)$ stands for Shannon's entropy. 
For some $A,X,Y$ this inequality is false, e.g., for constant $X,Y$ and non-constant $A$.
We provide a natural condition on the distribution of $A,X,Y$ implying inequality~\eqref{eq13-bis}. Then
we provide two combinatorial applications of the resulting conditional inequality and show that it implies the conditional
inequality from~\cite{kaced-romashchenko-2011}.

More specifically we consider the following condition:
\begin{equation}\label{cond-2-C} 
\begin{split}
&\text{for each quadruple  } a,a',x,y, \text{ if the probabilities }
\text{of all four events }\\
&[A=a,X=x],
[A=a,Y=y],
[A=a',X=x], 
[A=a',Y=y]\\
&\text{ are positive, }   
\text{then }a=a'.
\end{split}
\end{equation}

\begin{theorem}\label{thm-1}
The inequality~\eqref{eq13-bis}
holds for all random variables $A,X,Y$ satisfying~ \eqref{cond-2-C}.
\end{theorem}

We first prove this theorem and then show its combinatorial applications.

\section{Notation}

To simplify formulas,  we use the following  notation for the marginal distributions (conditional and unconditional): $p(a)$ denotes $\Pr[A=a]$,
$p(a,x)=\Pr[A=a,\ X=x]$, $p(a\mid x)=\Pr[A=a\mid X=x]$, 
$p(a,y)=\Pr[A=a,\ Y=y]$, and so on.

If $X$ is a random variable and $\cal E$ is an event in the same probabilistic space (and $\prob[{\cal E}]>0$), we denote by
 $X\mid \cal E$ the conditional distribution of $X$, i.e., the restriction of $X$ on the subspace corresponding to the event $\cal E$. For example, for 
 jointly distributed random variables $(X,Y)$ we denote by  $X\mid (Y=y)$  the conditional distribution of $X$ under the assumption $Y=y$.

\section{The proof of Theorem~\ref{thm-1}}

We apply the method of~\cite{zhang-conditional,zy98}. The crucial 
property of inequality~\eqref{eq13-bis} is that no term contains both $X$ and $Y$. 
The inequality~\eqref{eq13-bis} can be re-written in terms of unconditional entropies as follows:
\begin{equation}\nonumber %\label{eq8}
H(A,X)+H(A,Y)\le H(X)+H(Y)+H(A).
\end{equation}
Thus it means that 
the average value of the logarithm of the ratio
\begin{equation}\label{eq8}
\frac{p(x)p(y)p(a)}{p(a,x)p(a,y)}
\end{equation}
is less than or equal to $0$. The average is computed with respect 
to the distribution $p(a,x,y)$.  Computing the average, we take into 
account only the triples $(a,x,y)$ with 
positive probability. For such triples, both the numerator and denominator 
of ratio~\eqref{eq8} are positive and hence its logarithm is well 
defined. 

Now consider a new distribution $p'$ where 
$$
p'(a,x,y)=\begin{cases}
\frac{p(a,x)p(a,y)}{p(a)} & \text{if } p(a)>0,\\
0
& \text{otherwise.}
\end{cases} 
$$ 
Random variables distributed according to $p'$ can be
generated by the following process: First generate $a$
using the original distribution of $A$, then generate independently
$x$ using the conditional distribution $x \mid a$ and 
$y$ using the conditional distribution $y \mid a$. 

Notice that $p'(a,x,y)$ is positive if so is $p(a,x,y)$ but not the other way around.
However, ratio~\eqref{eq8} is still well defined and positive for all triples $a,x,y$ with positive $p'(a,x,y)$. 
Therefore we can compute the average value of the logarithm of~\eqref{eq8} using the distribution $p'$ in place of $p$. 
Moreover, changing the distribution does not affect the average.
%This follows from the equalities $p'(a,b,x) = p(a,b,x)$ and $p'(a,b,y) = p(a,b,y)$. 
Indeed, the logarithm of~\eqref{eq8} is the sum of logarithms of its factors.
Thus it suffices to show that the average of the logarithm of each factor is not
affected when $p$ is replaced by $p'$.
Let us prove this, say, for the factor $1/ p(a,x)$.

This factor does not depend on $y$.
Therefore the average  of its logarithm does not depend on how  $p(a,x)$ is split among   $p(a,x,y)$ for different values  $y$:
we just sum up $\log 1/ p(a,x)$ over all $a,x$ with weights $p(a,x)$. As $p(a,x)=p'(a,x)$, summing with weights $p'(a,x)$ will yield the same result.

By  Jensen's inequality\footnote{%
We need Jensen's inequality for the logarithmic function:
let $p_1,\dots,p_n$ be positive numbers that sum up to 1; then
$p_1\log x_1+\dots+p_n\log x_n\le \log(p_1x_1+\dots+p_nx_n)$.} 
the average value of the logarithm of the ratio
\eqref{eq8} with respect to the distribution $p'$ is
at most 
$$
\log\Bigl(\sum_{a,x,y: p'(a,x,y)>0}p(x)p(y)\Bigr).
$$

The condition~\eqref{cond-2-C} guarantees that 
for each $x,y$ there is at most one $a$ with
$p(a,x)>0,p(a,y)>0$ and hence 
$$
\log\Bigl(\sum_{a,x,y: p'(a,x,y)>0}p(x)p(y)\Bigr)\le
\log\Bigl(\sum_{x,y}p(x)p(y)\Bigr)=\log 1=0. 
$$

\section{Combinatorial applications of Theorem~\ref{thm-1}}

\subsection{A lower bound for the number of colors
  in edge colorings of bipartite graphs}

An \emph{edge coloring of a graph}
is an assignment of colors to its edges so that each two edges
sharing a node have different colors.
Finding the \emph{edge coloring number} (the minimum possible number of colors
in an edge coloring)
of a given graph is a classic problem of graph theory.
The study of edge coloring is motivated by theoretical aspects of graph theory 
as well as by numerous applications in information theory and
computer science (mostly by different types of scheduling problems,
see a survey in \cite{survey-edge-coloring}).

Vizing's theorem \cite{vizing} claims that the edge coloring number of a graph
is either its maximum degree
%of this graph
$d$ or $d+1$; for bipartite graphs the number of colors is always $d$.
%Thus, under the conditions of Corollary~\ref{prop1},  from Vizing's theorem it follows that every proper partition consists of at least $\max\{L,R\}$ edges.
From Theorem~\ref{thm-1} we can derive a much stronger
lower bound for edge colorings of bipartite graphs 
satisfying the following constraint:

\begin{definition}
  Call an edge coloring of
  a bipartite graph \emph{rich}
if for each pair 
 $$
 \langle \mbox{left vertex }x, \mbox{ right vertex }y\rangle 
 $$
 there is at most one color $a$ touching both $x$ and $y$
 (the latter means that there is an edge with color $a$ incident to $x$
 and an edge, maybe a different one, with color $a$ incident to $y$). 
\end{definition}

From Theorem~\ref{thm-1} we can derive the following
bound for rich colorings of bipartite graphs: 
\begin{corollary}\label{prop1}
Assume that the degree of each left vertex in a given bipartite
graph is at least $L$ and 
the degree of each right vertex is at least $R$.
Then the number of colors in every rich edge coloring of
the graph is at least $LR$.
\end{corollary}

\begin{IEEEproof}
%Let $M_1,\dots,M_K$ denote the given matchings.
Consider the uniform distribution on the set of edges of the graph.
Denote by $(A,X,Y)$ the following triple of jointly distributed random variables:

$X=[\mbox{the left end of the edge}]$, 

$Y= [\mbox{the right end of the edge}]$, 

$A= [\mbox{the color of the  edge}]$.

\noindent
As the coloring is rich,
%$M_1,\dots,M_K$ is a proper partition,
the triple $(A,X,Y)$ 
satisfies~\eqref{cond-2-C}: if 
both events $[A=a,X=x]$ and $[A=a,Y=y]$ have positive probabilities,
then both $x$ and $y$ are touched by 
$a$, and hence such $a$ is unique.
Therefore by Theorem~\ref{thm-1} we have
$H(A \mid X)+H(A \mid Y)\le H(A)$.% (see also Remark~\ref{rm4} on p.~\pageref{rm4}). 

\begin{comment}
Notice that for each color $a$ the probability $\prob[A=a]$ is proportional
to the number of edges with color $a$.
On the other hand, for each vertex $x$ touched by $a$,
the conditional probability $\prob[X=x\mid A=a]$ is inversely proportional
the number of edges with color $a$.
It follows that for every fixed vertex $x$ all colors
touching   $a$ are equiprobable.
In other words, conditional on $X=x$,
the value of $A$ is  uniformly distributed on the set of colors
touching $x$. Thus, $H(A \mid X)\ge \log L$. 
Similarly,  we have $H(A \mid Y)\ge \log R$. By Theorem~\ref{thm-1} we have
$H(A)\ge \log L+\log R$. It follows that the range of $A$ is at least $LR$.
\end{comment}

By construction, the distribution on the edges is uniform. Hence, for each
vertex $x$, the conditional distribution of edges incident to this $x$ is
also uniform.
All edges incident to one and the same vertex must have different colors.  
So for every fixed vertex $x$, all colors touching this $x$ are equiprobable.
In other words, conditional on $X=x$,
the value of $A$ is uniformly distributed on the set of colors
touching $x$. Thus, $H(A \mid X)\ge \log L$. 
Similarly,  we have $H(A \mid Y)\ge \log R$. By Theorem~\ref{thm-1} we have
$H(A)\ge \log L+\log R$. It follows that the range of $A$ is at least $LR$.
\end{IEEEproof}
\begin{remark}
In fact this proof gives a stronger result. 
Let us call by \emph{the left} and \emph{the right degrees} of an edge (in a bipartite graph) the degrees of its left and right ends.
Denote by $\tilde L$ and $\tilde R$ the geometric means of the left and the right degrees of the graph's edges. That is, if the degrees of the vertices in  the left part of the graph are $l_1,\ldots,l_n$ and  the degrees of the vertices in  right part  of the graph are $r_1,\ldots,r_m$ ($l_1+\ldots+l_n = r_1+\cdots+r_m = e$, where $e$ is the number of edges), then
 $$
 \tilde L := \big(l_1^{l_1}\cdots l_n^{l_n}\big)^{1/e}, \ 
  \tilde R := \big(r_1^{r_1}\cdots r_m^{r_m}\big)^{1/e}.
 $$
The proof of Corollary~\ref{prop1} explained above implies that  the number of colors in every rich edge coloring of the graph is at least $\tilde L \tilde R$.  Notice that $\tilde L \ge L$ and $\tilde R\ge R$ (these inequalities become equalities, if and only if the graph is uniform on the left or on the right respectively).
\end{remark}

In what follows we exhibit three examples of rich colorings. The first two examples are pretty trivial; in the third example,  Corollary~\ref{prop1} provides a non-trivial lower bound.

\begin{example}
For some bipartite graphs the lower bound $LR$ proven in Corollary~\ref{prop1} is tight.
Consider the simplest example: let $K_{R,L}$ be the complete bipartite graph with $
R$ left and $L$ right vertices so that the degree of each left
vertex is exactly $L$ and 
the degree of each right vertex is exactly $R$ (see  in Fig.~\ref{fig-2} an example for $R=3$ and $L=4$). This graph has $LR$ edges. We  may
color them into $LR$ colors, each edge having its unique
color.
This coloring is rich, as for each pair 
 $ \langle \mbox{left vertex }x, \mbox{ right vertex }y\rangle$
only the color of that edge touches both $x$ and $y$.
\begin{figure}[ht]
        \begin{center}
\includegraphics[scale=0.75]{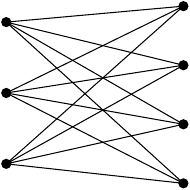}
        \end{center}
        \caption{Complete bipartite graph $K_{3,4}$.}\label{fig-2}
\end{figure}

For this  example it is easy to compute directly  
the minimum number of colors in a
rich coloring. Indeed,
no different edges
$(x_1,y_1)$, $(x_2,y_2)$ can share a color, as in that case
the pair $x_1,y_2$ would violate the condition: the color 
of the edge $(x_1,y_2)$ also touches both 
$x_1$ and $y_2$ and is different from the shared
color of $(x_1,y_1)$, $(x_2,y_2)$.
\end{example}

\begin{example}
In general, the lower bound  from Corollary~\ref{prop1} is not optimal.
To construct the simplest example, consider
the complete bipartite graph $K_{3,3}$ and delete from this graph three edges forming a perfect matching, see Fig.~\ref{fig-3}.
\begin{figure}[ht]
        \begin{center}
\includegraphics[scale=1]{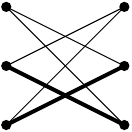}
        \end{center}
        \caption{Complete bipartite graph $K_{3,3}$ minus a perfect matching.}\label{fig-3}
\end{figure}
In this graph the degree of each vertex is $2$, so Corollary~\ref{prop1} claims
that every rich coloring has at least $2\cdot2=4$ colors.
However, it is easy to verify that the optimal rich coloring
has $5$ colors:  two edges (e.g., the pair of edges shown in bold in Fig.~\ref{fig-3}) may share a color, while each of the remaining edge must have a unique
color. 
\end{example}

The next example is less obvious and exhibits a series of rich
colorings for which Corollary~\ref{prop1} provides
a non-trivial lower bound.

\begin{example}
  \emph{Assume that a finite family $F$
    of pair-wise disjoint squares inside
    the square $[0;1)^2$ in the Euclidean plane is given,
      each square having the form $[a;b)\times [c,d)$.
Assume that all vertices of
those squares  have rational coordinates~\footnote{This
assumption is added for technical simplicity and may be dropped.}. 
  Assume further for each $x\in [0;1)$ there
  are at least $L$ squares in $F$ whose first projection
  includes $x$ and similarly for each $y\in [0;1)$ there
  are at least $R$ squares in $F$ whose second projection
  includes $y$ (see Fig.~\ref{fig}). Then $|F|\ge LR$.}

% \begin{wrapfigure}{r}{0.45\textwidth}
\begin{figure}[ht]
        \begin{center}
\includegraphics[scale=1.8]{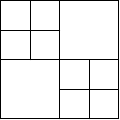}
        \end{center}
        \caption{The square  is partitioned into
          10 disjoint squares so that each vertical or horizontal
          line intersects 3
                squares.}\label{fig}
\end{figure}
%\end{wrapfigure}

  \begin{IEEEproof}
    Obviously, there is a natural $N$ such that
each of the given squares has the form 
$[i/N;j/N)\times [k/N;l/N)$ for integer $i,j,k,l\le N$.
Consider the graph whose left and right nodes
are rational numbers of the form $i/N$ with $0\le i< N$.
For each  given square $[a;b)\times [c,d)$
consider the diagonal $\{(a+t,c+t)\mid 0\le t< b-a=c-d\}$, see Fig~\ref{fig-5}.
\begin{figure}[ht]
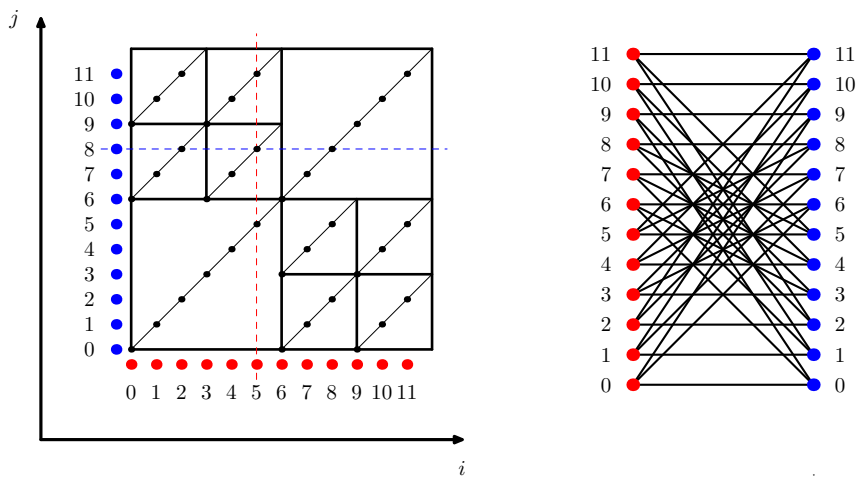

        \begin{center}
\includegraphics[scale=0.80]{fig-5.mps}\rule{1.5cm}{0cm}
\includegraphics[scale=0.80]{fig-6.mps}
        \end{center}
        \caption{Diagonals of the squares of the partition (points with rational coordinates $(i/N, j/N)$ for $N=12$ and the corresponding bipartite graph).}\label{fig-5}
\end{figure}
The edges of the graph are those pairs $(x,y)$ of nodes that
lie on such diagonals.
The edges of the resulting graph can be naturally colored into $|F|$
colors: the edges obtained from each diagonal  
are colored in a unique color. 
This coloring is rich: if the diagonal of a
square has a point of the form $(x,*)$
and a point of the form $(*,y)$, then that square
includes the point $(x,y)$ and there is at most
one such square (the squares are disjoint).
The left degree of the graph is at least $L$.
Indeed, for each $x_0\in [0;1)$ of the form $i/N$ there
are at least $L$ squares whose first projection
includes $x_0$ (see the dashed vertical in Fig~\ref{fig-5}). For each such
square $[a;b)\times [c,d)$ its diagonal intersects the vertical segment
    $\{(x_0,y)\mid 0\le y<1\}$, say, in the
    point $(x_0,y_0)$. Since both $a,c$ have the form $i/N$,
    the number $y_0=c+(x_0-a)$ also has this form and
    hence $(x_0,y_0)$ is an edge of the graph. Similarly,
    the right degree of the graph is at least $R$ (see the dashed horizontal line in Fig~\ref{fig-5}) and by Corollary~\ref{prop1}
    we have $|F|\ge LR$.
  \end{IEEEproof}  
  
\end{example}

\emph{Historical remark:}
The study of edge colorings with additional constraints is by all means not new, see, e.g.,  strong edge colorings \cite{strong-edge-coloring}, complete edge colorings \cite{complete-edge-coloring},  the Thue number of a graph \cite{alon-thue-number},  etc. Some versions of constraint edge coloring  have found direct applications in Information theory (e.g.,  \cite{icalp-constraint-edge-coloring,strong-edge-coloring-applications}).
The  problem concerned  in Corollary~\ref{prop1}  looks quite natural in the context of the variety  of edge coloring problems  investigated in graph theory, though we are not aware of any earlier studies or applications
of this specific variant of edge coloring.
%problem.

\begin{comment}
\begin{remark}
Counting matchings in bipartite graphs is a very interesting area of research with beautiful results. However, we believe that it
is irrelevant to our results, since we are only interested   
in minimum size of \emph{partitions in matchings} of edges of a graph.
It might happen that the number of matchings is huge while the minimal
size of a proper partition is small. E.g. for the complete bipartite
graph $K_{n,n}$ the former is more than $n!$,
and the latter is only $n^2$. On the other hand, for the graph $K_{1,n}$
both numbers are of the same order $O(n)$.
\end{remark}
\end{comment}
\subsection{A lower bound for the biclique cover  number of bipartite graphs}

\begin{definition}
For any bipartite graph $G=(V_1, V_2, E)$ (with the set of vertices $V_1\cup V_2$ and the set of edges $E\subset V_1\times V_2$) its \emph{biclique cover number} $bcc(G)$ is defined as the minimal number of bicliques (complete bipartite subgraphs) that cover all edges of $G$.
\end{definition}
Biclique coverings play an important role in communication complexity.  Specifically, the  \emph{non-deterministic communication complexity} (see \cite{kushilevitz-nisan}) of a predicate 
 $$
  P : U \times U \to \{0,1\}
 $$
can be defined as $\log bcc(G)$ for the bipartite graph $G=(V_1, V_2, E)$, where  $V_1=V_2=U$, and $E$ is the set of all pairs 
$(x,y)\in U\times U$ such that $P(x,y)=1$.

\begin{corollary}\label{thm-bcc}
Assume that the edges of a bipartite graph $G=(V_1, V_2, E)$ are colored in such a way that 
\begin{itemize}
\item[(*)] if edges  $(x,y')$ and $(x',y)$ of the graph have the same color $a$,
  and the vertices $x$ and $y$, as well as
  vertices $x'$ and $y'$, are also connected by edges,
 then the latter two edges also have color $a$.
\end{itemize} 
%\rom{\textbf{Question:} The quantifier over cliques ``... for every biclique $C$...'' look strange. Should we just say  that $x,x',y,y'$ are all connected by edges, i.e., they  induce a subgraph $K_{2,2}$?  A simpler (but weaker) version of the corollary requires only that $(x,y)$ are connected by an edge.}

Assume further that a probability distribution over the edges of the graph is given. Denote by $(X,Y,A)$ the random variables 
 where
 \begin{itemize}
 \item $X=[\mbox{the left end of the edge}]$, 
 \item $Y= [\mbox{the right end of the edge}]$, 
 \item $A= [\mbox{the color of the edge}]$.
 \end{itemize}
Then
$
bcc(G) \ge 2^{\frac12 (H(A \mid X) + H(A \mid Y)- H(A))}.
$
\end{corollary}

\begin{IEEEproof}
Assume that this graph $G$ can be covered by $t$ bicliques $C_1,\ldots, C_t$. Extend the distribution $(X,Y,A)$ and add another random variable: we define $Z$ as the index of a biclique $C_i$ that covers the edge $(X,Y)$. (If an edge belongs to several bicliques $C_i$, then we choose any of them.) Notice that $Z$ ranges over $\{1,\ldots, t\}$, so $H(Z)\le  \log t$.

The crucial point is that for a fixed value $i$ of $Z$  
the condition~(\ref{cond-2-C})  is satisfied. Indeed, assume 
that both $p(a,x|Z=i)$ and $p(a,y|Z=i)$ are positive. Then the biclique $C_i$ has edges $(x,y')$ and $(x',y)$, both with color $a$.
By property $(*)$  the color of the edge $(x,y)$ also equals $a$ and hence such $a$ is unique. By
Theorem~\ref{thm-1} for each conditional distribution $(A,X,Y) \mid Z=i$ the inequality~\eqref{eq13-bis} holds. Hence we get
$$%\begin{equation}\label{eq-bcc1}
H(A\mid X,Z) + H(A\mid Y,Z) \le H(A\mid Z).
$$%\end{equation}
It follows that 
$$%\begin{equation}\label{eq-bcc2}
H(A\mid X) -H(Z)+ H(A\mid Y) - H(Z) \le H(A).
$$%\end{equation}
Thus, we obtain
$
 t \ge 2^{H(Z)} \ge 2^{ \frac12[H(A \mid X ) + H(A \mid Y ) - H(A)]}.
$
\end{IEEEproof}

\begin{example}
Let us apply this corollary to a specific bipartite graph.
Consider the bipartite Kneser graph $KG_{n,k}=(V_1, V_2, E)$,
where both parts $V_1$ and $V_2$ consist of $k$-elements subsets of $\{1,\ldots,n\}$, and the set of edges $E\subset V_1\times V_2$ consists of all pairs of disjoint sets.
%,~\cite{kneser}.
Let us color the edge $(x,y)$  in color $x\cup y$ and consider the uniform probability distribution
over the edges of this graph. 
The condition $(*)$ is fulfilled. Indeed, assume we are given three pairs of disjoint $k$-element
subsets:  $(x,y)$, $(x,y')$ and $(x',y)$. Assume further that  
$x\cup y'=x'\cup y=a$.
Then $x=x'$ and $y=y'$ and hence  $x\cup y=a$ as well.
Hence 
$$
bcc(KG_{n,k})\ge  2^{ \frac12[H(A \mid X ) + H(A \mid Y ) - H(A)]}.
$$
We have ${n \choose 2k}$ equiprobable colors and hence 
$H(A)=\log_2 \binom{n}{2k}$. On the other hand,
$H(A|X)=H(A|Y)=\log_2 \binom{n-k}{k}$. Thus
$$
bcc(KG_{n,k})\ge \sqrt{\binom{n-k}{k}^2/\binom{n}{2k}}.
$$
If $n\gg k$ then $\binom{n-k}{k}^2/\binom{n}{2k}$ is close to $\binom{2k}k\approx 2^{2k}$ and we obtain a lower 
bound about $2^k$ for $bcc(KG_{n,k})$.
On the other hand, it is known that 
$
bcc(KG_{n,k}) \le 2^{O(k+\log \log n)}
$
(see \cite[Section~2.3]{kushilevitz-nisan}), so in the case $\Omega(\log \log n) \le k \ll n$ these lower and upper bounds are pretty close.

The proven bound in itself is of no interest; the simple and standard fooling set technique (see \cite{kushilevitz-nisan}) proves for this graph the bound 
$bcc(KG_{n,k})\ge \binom{2k}k$ that holds for all $n\ge 2k$. However, this simple example illustrates the connection between biclique cover and conditional information inequalities. It remains unknown whether a similar technique can surpass the fooling set method for other examples of graphs.
\end{example}

\emph{Historical remark:}
In graph theory  the minimum number of bicliques (complete bipartite subgraphs) needed to cover all edges of a  given graph is known as
\emph{the biclique cover number} or \emph{the bipartite dimension} of a graph. 
The problem of computing the bipartite dimension appears in different areas of computer science. In particular, the notions of bipartite partition and bipartite cover play the central role in communication complexity, \cite{kushilevitz-nisan}. 

The problem of determining the bipartite dimension is NP-hard even for bipartite graphs, \cite{biclique-np-hard}. A good  approximation or a nontrivial lower bound for the bipartite dimension of some particular classes of graphs may imply substantial progress in various problems of computational complexity, see \cite{kushilevitz-nisan,jukna-hraph-complexity,lee-survey,kulikov-yukna}. 
In Corollary~\ref{thm-bcc}  we proposed a new technique of lower bounds for the bipartite dimension. 
Establishing formal relations between our method and previous approaches to biclique cover  remains an open problem.

\section{A generalization of a conditional inequality from~\cite{kaced-romashchenko-2011}}

In this section we show that Theorem~\ref{thm-1} implies some conditional version of Ingleton's inequality for entropies.
So-called \emph{Ingleton's inequality} was originally formulated and  proven for ranks of linear subspaces, \cite{ingleton}. It turns out that a counterpart of this inequality reformulated in terms of  Shannon's entropy (for random variables) has many nontrivial applications. Though in general this inequality is not valid for entropies (see \cite{hammer}), it holds for distributions that satisfy some special properties (e.g., for random variables that enjoy the property of extracting the mutual information, or for variables with some properties of independence, see \cite{zhang-conditional,  matus-iii,  mmrv, matus,  dog}). In particular, 
in~\cite{kaced-romashchenko-2011} it was shown  that Ingleton's inequality for entropies holds for all distributions where the entropies satisfy some linear constraints:
\begin{theorem}[\cite{kaced-romashchenko-2011}] \label{thm-kr}
If random variables $X,Y,A,B$ satisfy the  the constraints
\begin{eqnarray}\label{eq-kr}
I(X\lon Y \mid A)=H(A \mid X,Y)=0,
\end{eqnarray}
then Ingleton's inequality 
\begin{eqnarray}\label{ingleton}
I(A\lon B)\le I(A\lon B \mid X)+I(A\lon B \mid Y) + I(X\lon Y)
\end{eqnarray}
holds for this distribution.
\end{theorem}
A noteworthy fact is  that  this result cannot be obtained as a direct implication of any unconditional linear inequality for Shannon's entropy. More precisely, whatever pair of  reals $\lambda_1, \lambda_2$ we take, the inequality
\begin{align*}
I(A\lon B)\le I(A\lon B \mid X)+I(A\lon B \mid Y) + I(X\lon Y) +\\ + \lambda_1 I(X\lon Y \mid A) + \lambda_2 H(A \mid X,Y)
\end{align*}
does not hold for some distribution, see~\cite{kaced-romashchenko-2013}. 

We claim that Ingleton's inequality
holds also under condition~\eqref{cond-2-C}, which is
weaker 
than~\eqref{eq-kr}. Moreover, even 
a stronger inequality than Ingleton's inequality
(namely the inequality~\eqref{eq-relativized} below),
holds under condition~\eqref{cond-2-C}.

\begin{theorem}\label{thm-3}
  (i) Ingleton's inequality~\eqref{ingleton} follows
  from the inequality 
\begin{eqnarray}\label{eq-relativized}
 H(A \mid X,B) + H(A \mid Y, B) \le H(A \mid B).
\end{eqnarray}

  (ii)
  Inequality~\eqref{eq-relativized} holds for all
  random variables $A,B,X,Y$ satisfying condition~\eqref{cond-2-C}.

  (iii)
  Condition~\eqref{cond-2-C} is implied by condition~\eqref{eq-kr}.
\end{theorem}

In brief, Theorem~\ref{thm-3}
states that \eqref{eq-kr} $\Rightarrow$ \eqref{cond-2-C} $\Rightarrow$
\eqref{eq-relativized} $\Rightarrow$ \eqref{ingleton}.  
The main novelty is the middle implication
\eqref{cond-2-C} $\Rightarrow$
\eqref{eq-relativized}, while the implications  \eqref{eq-kr} $\Rightarrow$ \eqref{cond-2-C} and \eqref{eq-relativized} $\Rightarrow$ \eqref{ingleton} are
almost straightforward (see the proof below)
and the implication \eqref{eq-kr} $\Rightarrow$ \eqref{ingleton}
was known (Theorem~\ref{thm-kr}).

\begin{IEEEproof}
(i) It is easy to verify that 
Ingleton's inequality~\eqref{ingleton} can be equivalently rewritten as
\begin{eqnarray}\label{ingleton-rewritten}
 H(A \mid X,B) + H(A \mid Y, B) \le H(A \mid B)+I(X\lon Y \mid A) + H(A\mid X,Y).
\end{eqnarray}
and hence follows from~\eqref{eq-relativized}.
Notice that under the constraints~\eqref{eq-kr},
Ingleton's inequality is equivalent to~\eqref{eq-relativized}.

(ii)
Note that  inequality~\eqref{eq-relativized} is
a relativized version of~\eqref{eq13-bis}
(the word \emph{relativization} here means that we insert a new 
condition in all entropy expressions). This  similarity between inequalities~\eqref{eq13-bis}  and~\eqref{eq-relativized} suggests that  
Theorem~\ref{thm-kr}  can be deduced from Theorem~\ref{thm-1}.
The key observation is  that  condition~\eqref{cond-2-C} is ``relativizable'':   
property \eqref{cond-2-C} remains true if we restrict the initial probabilistic space to some subspace.
\begin{lemma}%[proved in Appendix]
\label{lemma-3}
If a tuple of random variables $(A,X,Y)$ satisfies \eqref{cond-2-C}, then for each event $\mathcal E$ having positive probability the conditional random
variables of $(A,X,Y)\mid \mathcal E$ satisfy~\eqref{cond-2-C}.
\end{lemma}
\begin{IEEEproof}
Assume that 
the four probabilities
\begin{align*}
\prob[X=x,\   A=a\ \mid \mathcal E], 
\prob[Y=y,\  A=a\ \mid  \mathcal E], \\
\prob[X=x,\   A=a'\mid  \mathcal E], 
\prob[Y=y,\  A=a'\mid  \mathcal E]
\end{align*}
are positive. Then the unconditional probability
of each of these events is positive as well and hence $a=a'$ by~\eqref{cond-2-C}.
\end{IEEEproof}

Now we can show that  \eqref{eq-relativized} follows from condition~\eqref{cond-2-C}.
Indeed, for every possible value $b$ of $B$  
the lemma
 %~\ref{lemma-3}
guarantees that~\eqref{cond-2-C} remains valid
conditional on the event $B=b$. 
By Theorem~\ref{thm-1} this implies 
$$
 H(A \mid X, B=b) + H(A \mid Y, B=b) \le H(A \mid B=b),
$$
and taking the average over all values $b$ we get~\eqref{eq-relativized}. 
%This claim can be formulated as the following corollary of Theorem~\ref{thm-1} (generalizing Theorem~\ref{thm-kr} in the form~\eqref{eq21}).

(iii)
Inequality $I(X\lon Y \mid A)=0$ means that
$$
p(a,x,y)p(a)=p(a,x)p(a,y)
$$ for all triples $a,x,y$. Thus it 
implies that for each triple $a,x,y$ of values of $A,X,Y$, if both probabilities $p(a,x)$ and $p(a,y)$ are positive, then $p(a,x,y)$ is also positive. 
Hence, if for  some $a, a', x, y$ all the four probabilities 
$$
p(a,x),
p(a,y),
p(a',x), 
p(a',y)
$$
are positive (the assumption of \eqref{cond-2-C}), then it follows that the probabilities 
$
p(a,x,y)
$
and
$
p(a',x,y)
$
must be also positive.

Now we employ the condition $H(A|X,Y)=0$ (which means that the value of $A$ is a deterministic function of $(X,Y)$). If both probabilities 
$
p(a,x,y)
$
and 
$
p(a',x,y)
$
are positive,  then $a=a'$, and we obtain the conclusion of~\eqref{cond-2-C}.

\end{IEEEproof}

Note that in general inequality~\eqref{eq-relativized}
is stronger than Ingleton's inequality~\eqref{ingleton}.
For instance, let $B$ be constant, let $X,Y$ be 
independent uniformly distributed random bits, and let $A=X\oplus Y$.
Then inequality~\eqref{eq-relativized} specializes to
$1+1\le 1$ and hence is wrong, while Ingleton's inequality~\eqref{ingleton}
specializes to $0\le 0+0+0$ (or to $1+1\le 1+1+0$,
if written in the form~\eqref{ingleton-rewritten}) and hence is true.

Note also that in general condition~\eqref{cond-2-C}
is weaker than condition~\eqref{eq-kr}.
For instance, let $A$ be constant and let $X,Y$ be any dependent
random variables.
%Then condition~\eqref{eq-kr} does not hold
%while~\eqref{eq-kr} trivially holds.

\section*{Acknowledgment}

The authors  are grateful to Maxim Popov for valuable comments on a preliminary version of this paper. The authors thank anonymous referees of IEEE Trans. Inform. Theory for many comments and suggestions.

\end{document}